\documentclass[twocolumn,aps,tightenlines]{revtex4}
\usepackage[latin9]{inputenc}
\usepackage{amsmath}
\usepackage{amssymb}
\usepackage{graphicx}
\usepackage{esint}

\makeatletter
\@ifundefined{textcolor}{}
{%
 \definecolor{BLACK}{gray}{0}
 \definecolor{WHITE}{gray}{1}
 \definecolor{RED}{rgb}{1,0,0}
 \definecolor{GREEN}{rgb}{0,1,0}
 \definecolor{BLUE}{rgb}{0,0,1}
 \definecolor{CYAN}{cmyk}{1,0,0,0}
 \definecolor{MAGENTA}{cmyk}{0,1,0,0}
 \definecolor{YELLOW}{cmyk}{0,0,1,0}
}

\usepackage{amsfonts}
\usepackage{mathrsfs}\usepackage{bbm}\usepackage{bbold}\usepackage{subfigure}
\newcommand{\rmd}{{\rm d}}

\newcommand{\bk}{{\bf k}}

\newcommand{\fvec}[1]{\boldsymbol{#1}}

\newcommand{\half}{\frac{1}{2}}

\makeatother

\begin{document}

\title{Interplay between tetragonal magnetic order, stripe magnetism, and
superconductivity in iron-based materials}

\author{Jian Kang}

\affiliation{School of Physics and Astronomy, University of Minnesota, Minneapolis,
MN 55455, USA}

\author{Xiaoyu Wang}

\affiliation{School of Physics and Astronomy, University of Minnesota, Minneapolis,
MN 55455, USA}

\author{Andrey V. Chubukov}

\affiliation{School of Physics and Astronomy, University of Minnesota, Minneapolis,
MN 55455, USA}

\author{Rafael M. Fernandes}

\affiliation{School of Physics and Astronomy, University of Minnesota, Minneapolis,
MN 55455, USA}
\begin{abstract}
Motivated by recent experiments in Ba$_{1-x}$K$_{x}$Fe$_{2}$As$_{2}$
{[}A. E. B\"ohmer \textit{et al}, to be published{]}, we analyze the type
of spin-density wave (SDW) order in doped iron-pnictides and the discontinuities
of the superconducting transition temperature $T_{c}$ in the coexistence
phase with SDW magnetism. By tracking the magnetic transition line
$T_{N}\left(x\right)$ towards optimal doping within an itinerant
fermionic model, we find a sequence of transitions from the stripe-orthorhombic
($C_{2}$) SDW order to the tetragonal ($C_{4}$) order and then back
to the $C_{2}$ order. We argue that the superconducting $T_{c}$
has two discontinuities -- it jumps to a smaller value upon entering
the coexistence region with the $C_{4}$ magnetic phase, and then
jumps to a larger value inside the SDW state when it crosses the boundary
between the $C_{4}$ and $C_{2}$ SDW orders. The full agreement with
the experimental phase diagram provides a strong indication that the
itinerant approach is adequate to describe the physics of weakly/moderately
doped iron-pnictides.
\end{abstract}
\maketitle
\textit{Introduction.}~~~One of the key features of the iron-based
superconductors is the proximity or coexistence of superconductivity
with a spin-density wave (SDW) magnetic order~\cite{review}. The
stripe-type magnetic order (spins aligned ferromagnetically in one
direction and antiferromagnetically in the other) has been observed
experimentally in numerous undoped and weakly doped materials below
$T_{N}\sim150K$ \cite{stripe,Dagotto12}. Such an order breaks the
$O(3)$ spin-rotational symmetry and also breaks the tetragonal ($C_{4}$)
lattice rotational symmetry down to orthorhombic ($C_{2}$). Theoretically,
the stripe order has been found in both itinerant~\cite{Eremin10,Fernandes12,igor_m,zlatko,Brydon11,vafek}
and localized spin~\cite{localized,Lorenzana08,Batista11} approaches
to Fe-pnictides. In the itinerant scenario, stripe order originates
from the interaction between fermions near hole and electron pockets,
which are separated by ${\bf Q}_{1}=(0,\pi)$ and ${\bf Q}_{2}=(\pi,0)$
in the Fe-only Brillouin zone. Fluctuations of the stripe order above
the magnetic ordering temperature $T_{N}$ were further argued~\cite{Fernandes12,Fernandes14}
to split the $O(3)$ and $C_{4}$ transitions and give rise to the
observed nematic-type order at $T_{N}<T<T_{\mathrm{nem}}$ in which
the $C_{4}$ lattice rotational symmetry is broken down to $C_{2}$
but the $O(3)$ spin-rotational symmetry remains unbroken.

\begin{figure}[htbp]
\centering{}\includegraphics[width=0.49\columnwidth]{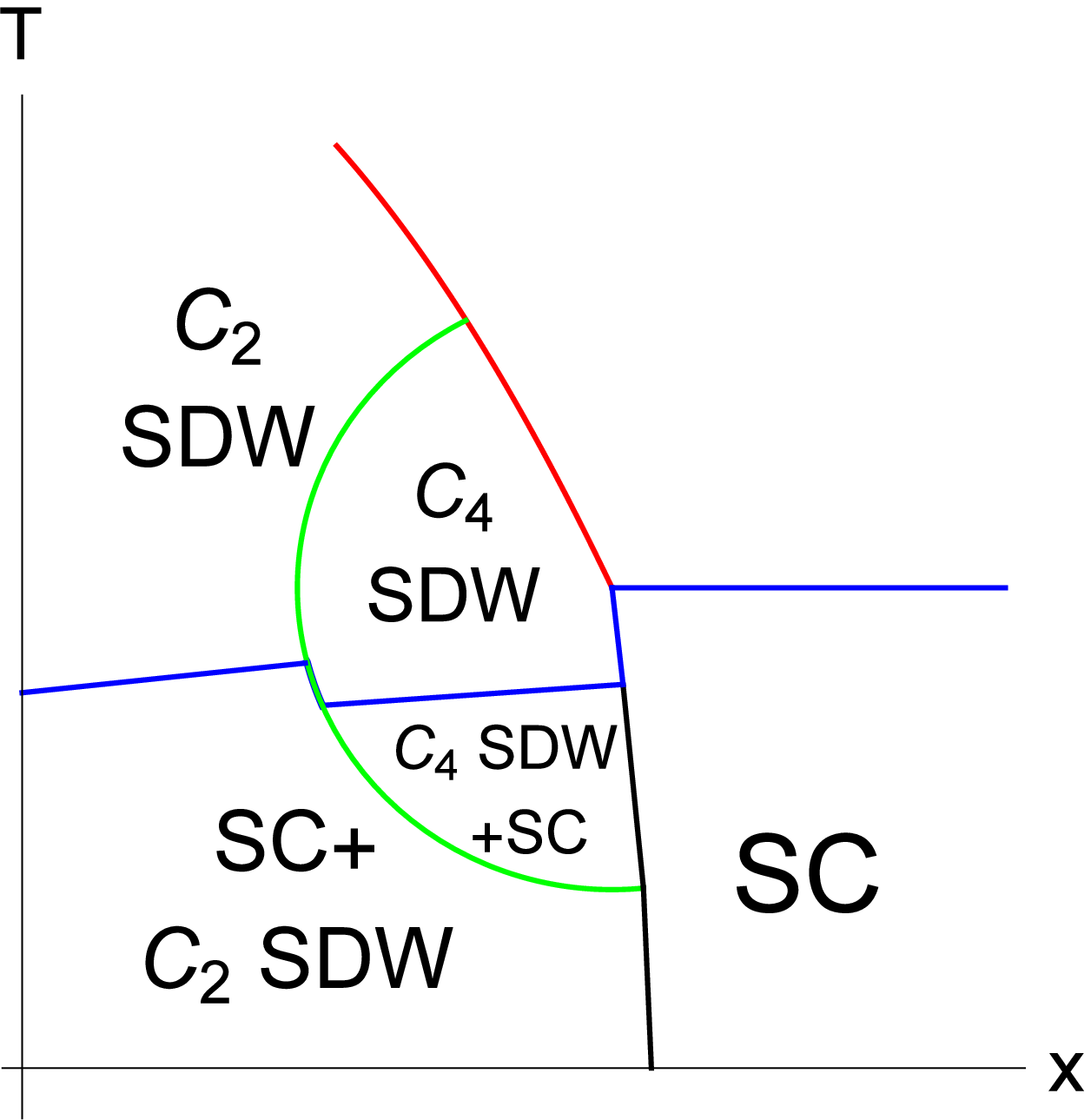}\hfill{}
\includegraphics[width=0.49\columnwidth]{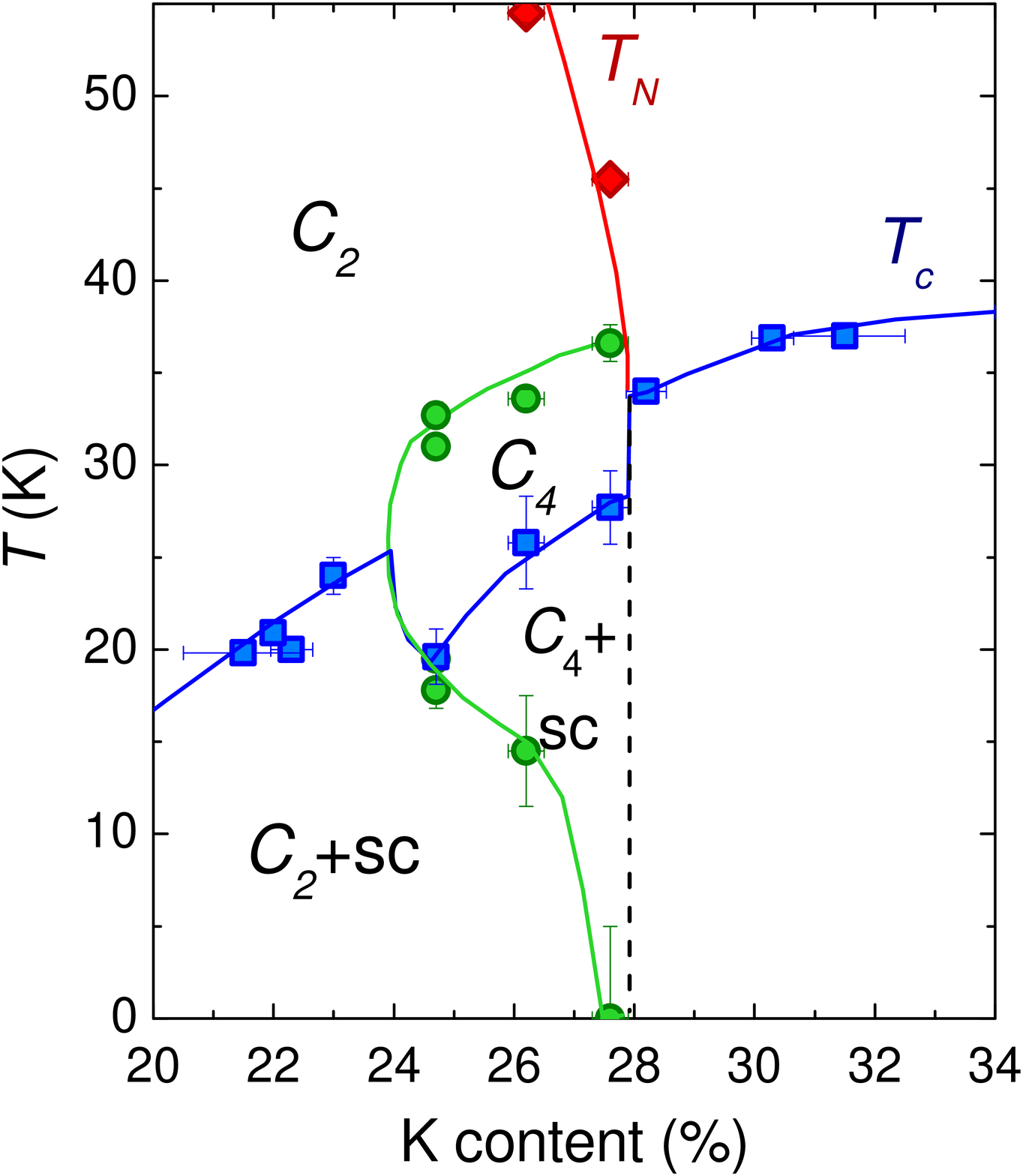}
\protect\protect\caption{Schematic phase diagram resulting from our itinerant model (left)
and experimental phase diagram of Ref. \cite{anna} for Ba$_{1-x}$K$_{x}$Fe$_{2}$As$_{2}$
(right). Blue lines refer to the (second-order) SC phase transition
whereas the green and red lines refer to the (first-order) $C_{4}-C_2$
and normal state-SDW phase transitions, respectively. Black lines refer to the first order SDW phase transitions inside SC phase. The four experimental
features discussed in the main text (i)-(iv) are naturally captured
by the itinerant model. The precise shapes of the transition lines
is non-universal and depends on details of the model not included
here. \label{Fig:SCSDWPhase}}
\end{figure}

A recent experiment on the hole-doped 122 Fe-pnictide Ba$_{1-x}$K$_{x}$Fe$_{2}$As$_{2}$
(Ref. \cite{anna}), however, found that the stripe magnetic configuration
does not persist at all dopings where magnetic order has been observed.
Instead, in some doping range, the stripe magnetic phase is replaced
by another SDW state in which the tetragonal $C_{4}$ symmetry is
unbroken. Neutron scattering experiments in the related compound Ba$_{1-x}$Na$_{x}$Fe$_{2}$As$_{2}$
(Ref. \cite{ray}) reported a similar $C_{4}$ SDW phase, with the
spin response still peaked at ${\bf Q}_{1}$ and ${\bf Q}_{2}$. The
most natural explanation for such a $C_{4}$ SDW phase is a magnetic
configuration with equal magnetic spectral weight at the ${\bf Q}_{1}$
and ${\bf Q}_{2}$ ordering vectors, resulting either in a orthogonal
checkerboard or in a non-uniform spin pattern (see Refs. \cite{Lorenzana08,Eremin10,Berg10,Brydon11,Giovannetti11,Wang14,Wang_arxiv_14}
). Hereafter we label this phase as $C_{4}$ SDW order and the stripe
phase as $C_{2}$ SDW order. Both $C_{4}$ and $C_{2}$ SDW orders
were found experimentally~\cite{ray,anna} to coexist with superconducting
(SC) order. The detailed analysis of the boundaries of the $C_{4}$
SDW phase in the phase diagram of Ba$_{1-x}$K$_{x}$Fe$_{2}$As$_{2}$
(Ref. \cite{anna}) shows several prominent features that require
theoretical explanation (see Fig. \ref{Fig:SCSDWPhase}): (i) The
$C_{4}$ SDW phase is confined to a narrow doping range with relatively
low $T_{N}$ values, being sandwiched by two regions with $C_{2}$
SDW order at lower and higher doping levels. (ii) The $C_{4}$ SDW
phase is confined to the near-vicinity of the magnetic instability
line $T_{N}$ and does not extend deep into the magnetically ordered
region. (iii) The superconducting $T_{c}$ is discontinuous at the
onset of the coexistence with $C_{4}$ SDW, where it jumps \textit{down}
by a finite amount. (iv) $T_{c}$ is again discontinuous when it crosses
the boundary between $C_{4}$ and $C_{2}$ SDW orders inside the SC
coexistence region, jumping \textit{up} by a finite amount.

In this communication we argue that all four features can be naturally
explained within the itinerant scenario for magnetism in iron-pnictides.
We depart from a model of interacting electrons located near hole
and electron pockets and derive and analyze the Ginzburg-Landau (GL)
free-energy for the coupled SDW and SC order parameters. We first
analyze the structure of the SDW order alone. We argue, based on the
analysis of GL expansion to fourth order, that the parameter that
determines whether the SDW order is $C_{2}$ or $C_{4}$ immediately
below $T_{N}$ changes sign twice along the $T_{N}(x)$ line. For
large and small values of $T_{N}$ the stripe order wins, whereas
for intermediate $T_{N}$ values the $C_{4}$ SDW order wins, explaining
observation (i) above. We then extend the GL analysis into the ordered
phase by expanding it to higher (sixth) order, showing that larger
values of the magnetic order parameter favor the stripe $C_{2}$ phase,
even if the initial instability is towards the $C_{4}$ SDW order.
This restricts the $C_{4}$ phase to the vicinity of the $T_{N}$
instability line, explaining the experimental feature (ii).

We then analyze the GL model for interacting SDW and $s^{+-}$ SC
order parameters. We first argue that the jump of $T_{c}$ to a smaller
value at the onset of coexistence with SDW is a natural consequence
of the experimental fact that the SC transition line $T_{c}$ crosses
$T_{N}$ at doping levels where the magnetic transition is first-order.
Specifically, the sign of the biquadratic coupling between the $s^{+-}$
SC and SDW order parameters \cite{Vorontsov09,fern_schm} is such
that the jump in the SDW order parameter at the point where the $T_{N}$
and $T_{c}$ lines meet causes a jump of $T_{c}$ to a smaller value,
consistent with observation (iii). We then analyze the behavior of
$T_{c}$ inside the SDW+SC coexistence state, as it crosses the boundary
between the $C_{4}$ and $C_{2}$ SDW phases. We argue that $T_{c}$
again jumps, this time to a larger value. The discontinuity is due
to the fact that the energy of the $C_{2}+\mathrm{SC}$ phase is lower
than that of the $C_{4}+\mathrm{SC}$ phase by a finite amount because
in the $C_{2}+\mathrm{SC}$ phase the system necessarily develops
a $d$-wave component of the SC order parameter due to the breaking
of the $C_{4}$ symmetry \cite{Fernandes_Millis}. We show that this
gives rise to an additional gain of condensation energy, resulting
in a higher $T_{c}$ in the $C_{2}+\mathrm{SC}$ phase compared to
$T_{c}$ in the $C_{4}+\mathrm{SC}$ phase. This is consistent with
the experimental observation (iv). This last effect is additionally
enhanced in Ba$_{1-x}$K$_{x}$Fe$_{2}$As$_{2}$ because the sub-leading
$d$-wave instability is nearly degenerate with the leading $s^{+-}$
instability \cite{Kuroki09,GraserSDDeg,s_plus_id_Thomale,Maiti11,Kang14},
as seen for instance by recent Raman experiments \cite{raman_mode1,raman_mode2}.

We interpret the good agreement between our itinerant theory and the
experimental data, including fine details, as a strong indication
that the itinerant approach to magnetism in Fe-pnictides is capable
to explain the physics of these materials. The situation may be different
in 11 Fe-chalcogenides where, at least for the parent FeTe, magnetic
order involves different momenta and cannot be naturally obtained
within an itinerant scenario~\cite{igor_m,yin10,Dagotto12,fete}.

\textit{The model.}~~~We consider the three-band 2D model with
one circular hole pocket centered at $(0,0)$ and two elliptical electron
pockets centered at $(\pi,0)$ and $(0,\pi)$ in the Fe-only Brillouin
zone. This is the minimal model to account for itinerant ${\bf Q}_{1}$/${\bf Q}_{2}$
magnetism \cite{Vorontsov09,fern_schm,Eremin10,Fernandes12}. The
inclusion of other two hole pockets complicates calculations but does
not lead to new physics. We follow previous works~\cite{Eremin10,Fernandes12}
and approximate the band dispersions as parabolic ones, $H_{0}=\sum_{\fvec ka\alpha}\epsilon_{\fvec ka}c_{\fvec ka\alpha}^{\dagger}c_{\fvec ka\alpha}$,
with:
\begin{equation}
\begin{aligned}\epsilon_{\fvec kh} & =-\epsilon_{\fvec k}=\frac{k^{2}}{2m}-\epsilon_{0}\\
\epsilon_{e_{1/2},\fvec k+\fvec Q_{1/2}} & =\epsilon_{\fvec k}+\left(\delta_{\mu}\pm\delta_{m}\cos2\theta\right)
\end{aligned}
\label{Eqn:BandStr}
\end{equation}
where $\fvec k$, $a$, and $\alpha$ refer to the momentum, band,
and spin indices, respectively. $\delta_{\mu}$ measures the chemical
doping, $\delta_{m}$ accounts for the ellipticity of the electron
pockets, and $\theta$ is the angle around an elliptical electron
pocket. The two interactions relevant to SDW order are density-density
($U_{1}$) and pair-hopping ($U_{3}$) interactions between hole and
electron pockets (see Refs. \cite{Eremin10,Fernandes12}). They act
identically in the SDW channel and drive the system towards the SDW
state with ordering vectors $\fvec Q_{1}$/$\fvec Q_{2}$, which are
the momentum displacements between the centers of electron and hole
pockets. To obtain the GL free-energy we introduce two SDW fields
$\fvec M_{i}(\fvec q)=\left(U_{1}+U_{3}\right)\sum_{k}c_{\fvec k+\fvec q,h}^{\dagger}\frac{\fvec\sigma}{2}c_{\fvec k+\fvec Q_{i},e_{i}},$
apply a Hubbard-Stratonovich transformation to decouple the 4-fermion
interaction, integrate out the fermions, and expand the free-energy
in powers of the SDW fields. To sixth order in $\fvec M_{i}$, the
free energy is expressed as
\begin{equation}
\begin{aligned}F(\fvec M_{i}) & =\frac{a}{2}\left(\fvec M_{1}^{2}+\fvec M_{2}^{2}\right)+\frac{u}{4}\left(\fvec M_{1}^{2}+\fvec M_{2}^{2}\right)^{2}\\
 & -\frac{g}{4}\left(\fvec M_{1}^{2}-\fvec M_{2}^{2}\right)^{2}+w\left(\fvec M_{1}\cdot\fvec M_{2}\right)^{2}\\
 & -\frac{v}{6}\left(\fvec M_{1}^{2}-\fvec M_{2}^{2}\right)^{2}\left(\fvec M_{1}^{2}+\fvec M_{2}^{2}\right)\\
 & +\frac{\gamma}{6}\left(\fvec M_{1}^{2}+\fvec M_{2}^{2}\right)^{3}+{\tilde{F}}(\fvec M_{i})
\end{aligned}
\label{Eqn:FMag}
\end{equation}
where ${\tilde{F}}(\fvec M_{i})$ stands for the terms with spatial
and time derivatives. Note that the free-energy itself is invariant
under $C_{4}$ rotations. All coefficients in Eq. (\ref{Eqn:FMag})
are the convolutions of fermionic Green's functions (two for $a$,
four for $u,g$ and $w$, and six for $v$ and $\gamma$), and are
expressed in terms of the band parameters from Eq.~\ref{Eqn:BandStr}.
We present the explicit results for these coefficients in the Supplementary
Information (SI). All coefficients except for $w$, which vanishes
in our model, are non-zero and depend on doping, temperature, and
degree of ellipticity of the electron pockets.

\textit{$C_{2}$ vs $C_{4}$ magnetism.}~~~ In the mean-field approximation,
one neglects ${\tilde{F}}(\fvec M_{i})$ and obtains the equilibrium
values of ${\bf M}_{1}$ and ${\bf M}_{2}$ by minimizing the free-energy.
Let us first assume that ${\bf M}_{1,2}$ are small and restrict the
analysis up to the quartic terms -- i.e. we approach the transition
from the paramagnetic side and check what happens immediately below
$T_{N}$. One then finds in a straightforward way that the system
develops $C_{2}$ order when $g>0$ and $C_{4}$ order when $g<0$.
For $C_{2}$ order, either $\fvec M_{1}$ or $\fvec M_{2}$ vanishes,
whereas for $C_{4}$ order, $\fvec M_{1}^{2}=\fvec M_{2}^{2}$. For
$g>0$, fluctuations contained in ${\tilde{F}}(\fvec M_{i})$ give
rise to an intermediate nematic phase in which the $C_{4}$ symmetry
is broken to $C_{2}$, but $\left\langle \fvec M_{i}\right\rangle =0$.

\begin{figure}[htbp]
\centering \subfigure[\label{Fig:SDW:Phase}]{\includegraphics[scale=0.25]{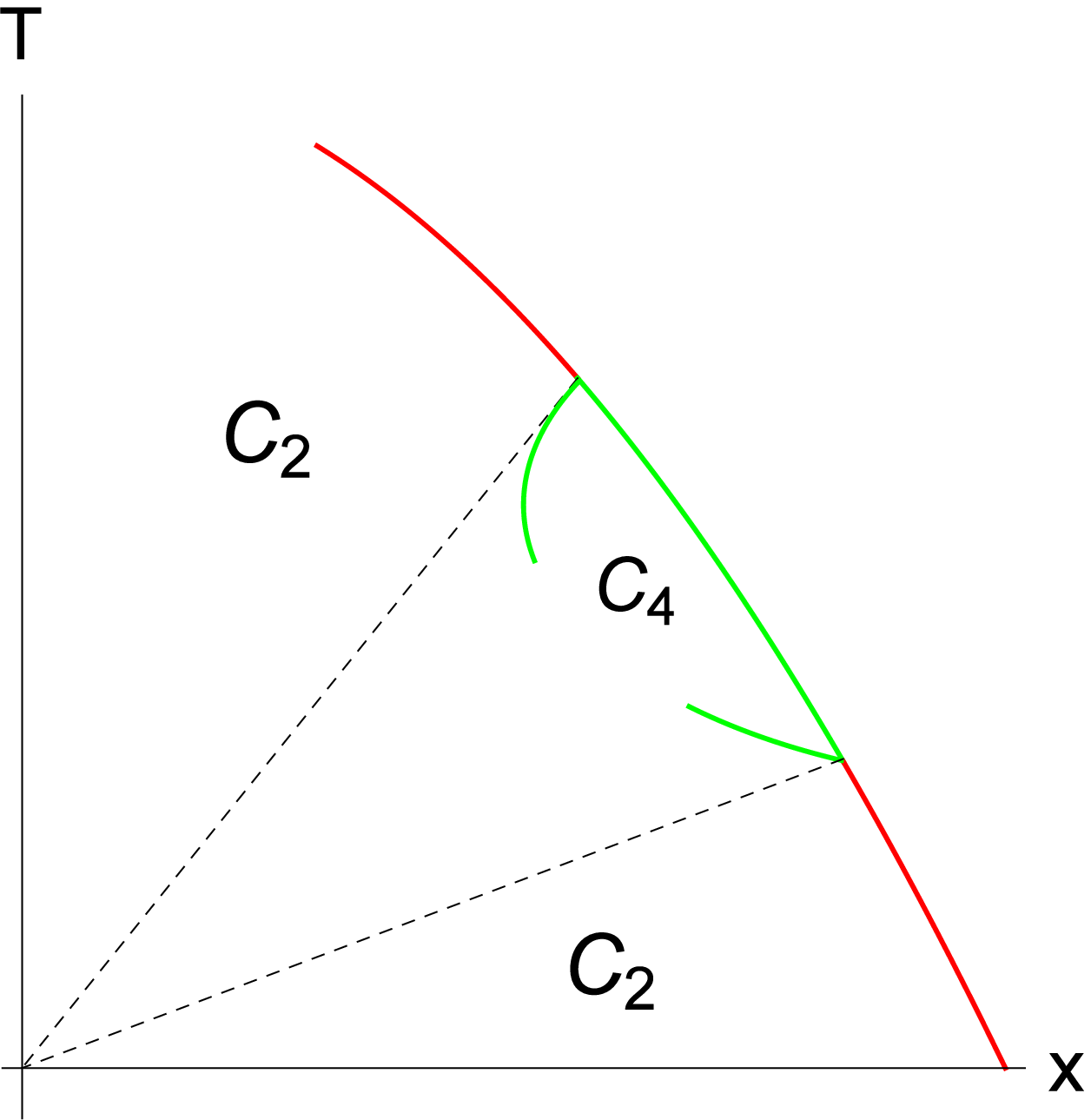}}
\subfigure[\label{Fig:SDW:gvCoe}]{\includegraphics[scale=0.4]{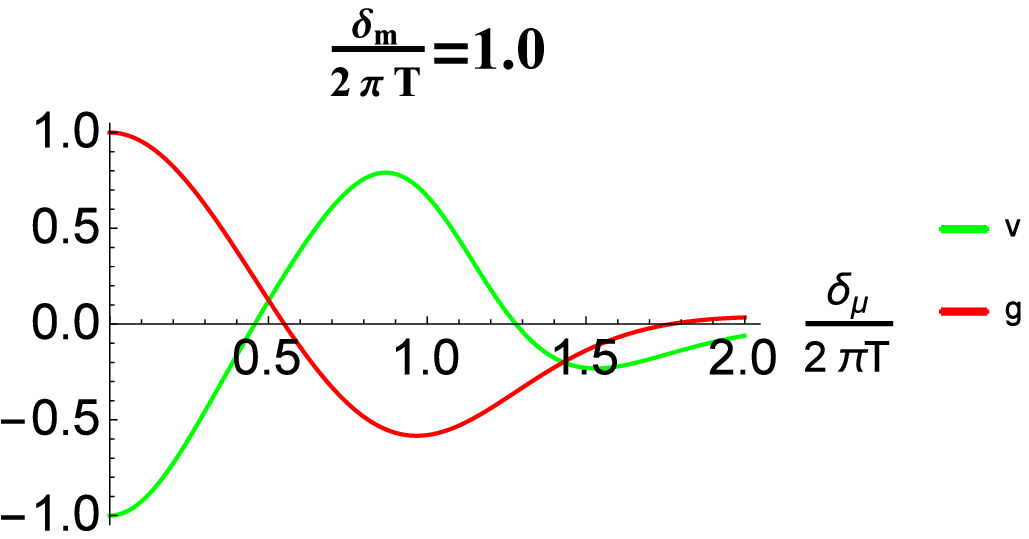}}
\protect\protect\caption{(a) The schematic phase diagram when SC is not included. The dashed
(solid) lines give the phase boundary of $C_{4}$ SDW order in the
absence (presence) of higher-order terms in the free energy. (b) The
coefficients $g$ and $v$ in Eq.~\eqref{Eqn:FMag} (normalized to
their $\delta_{\mu}=0$ values) as functions of $\delta_{\mu}/(2\pi T)$
for $\delta_{m}/(2\pi T)=1$. Note that $g$ and $v$ change sign
twice as $\delta_{\mu}/(2\pi T)$ becomes larger. This behavior is
generic for other values of $\delta_{m}/(2\pi T)$ (see the SI).}

\label{Fig:SDW}
\end{figure}

In Fig.~\ref{Fig:SDW} we show the behavior of $g$ as a function
of $\delta_{\mu}/T_{N}$, where the chemical potential $\delta_{\mu}$
is proportional to doping. For simplicity we show a plot for fixed
$\delta_{m}/T_{N}$, but the behavior described here is generic (see
SI). In the limit of high transition temperatures $T_{N}$ (small
doping), we find that $g\approx\frac{31\zeta(5)N_{f}\delta_{m}^{2}}{64\pi^{4}T_{N}^{4}}>0$,
whereas as $T_{N}\rightarrow0$ (optimal doping), $g\approx\frac{\delta_{m}^{2}N_{f}}{|\delta_{\mu}|(\delta_{\mu}^{2}-\delta_{m}^{2})^{3/2}}>0$.
For this last result, we used the fact that $|\delta_{\mu}|>\delta_{m}$
when $T_{N}\rightarrow0$ \cite{Vorontsov09}. Thus, in the high and
low $T_{N}$ regimes, the system develops a stripe $C_{2}$ order
below $T_{N}(x)$. However, at intermediate $T_{N}$ we find that
$g$ necessarily changes sign and becomes negative over some range
of doping concetrations. Once $g<0$, the system develops $C_{4}$
order. This explains the experimental observation (i) in Ba$_{1-x}$K$_{x}$Fe$_{2}$As$_{2}$,
i.e. that the doping range with $C_{4}$ order is sandwiched between
two doping regions with $C_{2}$ order. Note that since $g$ remains
very small after it changes sign for the second time, the energies
of both the $C_{4}$ and the $C_{2}$ SDW states are very close \cite{ray}.

We next analyze how the boundaries of the $C_{4}$ order evolve as
the system moves into the SDW phase. If we would restrict our analysis
to the fourth-order terms in the free energy, the $C_{4}$ phase would
extend all the way down to $T=0$ (dashed lines in Fig. \ref{Fig:SDW}a).
However, once the SDW order develops, higher-order terms in the GL
free energy become relevant. In particular, the sixth-order term relevant
for the $C_{4}-C_{2}$ transition in Eq. (\ref{Eqn:FMag}) is $-\frac{v}{6}\left(\fvec M_{1}^{2}-\fvec M_{2}^{2}\right)^{2}\left(\fvec M_{1}^{2}+\fvec M_{2}^{2}\right)$.
This term has the same $\left(\fvec M_{1}^{2}-\fvec M_{2}^{2}\right)^{2}$
structure as the fourth order term $-\frac{g}{4}\left(\fvec M_{1}^{2}-\fvec M_{2}^{2}\right)^{2}$
but scales additionally with the magnitude of ${\bf M}^{2}$. Combining
the sixth-order and the fourth-order terms we find that the location
of the boundaries between the $C_{4}$ and $C_{2}$ phases inside
the SDW-ordered region is determined by the zeros of
\begin{equation}
\tilde{g}=g+\frac{2}{3}v\left(\fvec M_{1}^{2}+\fvec M_{2}^{2}\right)\ ,\label{Eqn:EffG}
\end{equation}
The analytic expression for the coefficient $v$ is presented in the
SI. $v$ can by itself be positive or negative, depending on doping.
We show how the sign of $v$ changes as function of $\delta_{\mu}$
in Fig. \ref{Fig:SDW}. We note that in most of the region where $g<0$,
the coefficient $v$ is positive, hence the sixth-order term prefers
the $C_{2}$ phase and progressively shrinks the temperature range
with $C_{4}$ order as the SDW order grows, resulting in the boundaries
of the $C_{4}$ phase shown by the solid lines in Fig. \ref{Fig:SDW}a
.We see this behavior as a strong indication that the $C_{4}$ phase
progressively yields to the $C_{2}$ phase as SDW order grows, in
agreement with the experimental determination of the $C_{4}$ line
in Ba$_{1-x}$K$_{x}$Fe$_{2}$As$_{2}$ (feature (ii) discussed in
the Introduction).

\textit{The interplay between SDW and superconductivity.}~~~ We
now consider how the existence of both $C_{2}$ and $C_{4}$ phases
affects the behavior of $T_{c}$ in the state where SC and SDW coexist
microscopically \cite{coexist_Kdoped_1,coexist_Kdoped_2}. We assume,
as many authors before us, that superconductivity outside the coexistence
region with SDW is of $s^{+-}$ type, i.e. the SC order parameter
is $s-$wave, but changes sign between hole and electron pockets \cite{reviews_pairing}.
We also assume that the SC transition in the absence of SDW is second
order. Two inputs are needed to proceed our analysis: the character
of the SDW transition and the location of the crossing point between
the $T_{c}(x)$ and $T_{N}(x)$ transition lines.

The $C_{2}-C_{4}$ transition is obviously first-order since in the
$C_{2}$ phase one of the magnetic order parameters ${\bf M}_{1,2}$
is zero while in the $C_{4}$ phase both have equal magnitude. The
character of the transition from the paramagnetic to the $C_{4}$
(or $C_{2}$) phase is determined by the interplay between the $u$
and the $\gamma$ terms in the GL free energy of Eq. (\ref{Eqn:FMag})
(Refs. \cite{Vorontsov09,Fernandes12}). We found (see SI) that at
least in the portion of the phase diagram where $g$ is negative,
$u$ is also negative and $\gamma$ is positive, implying that the
transition into the $C_{4}$ SDW phase is first-order \cite{comm}.
This is consistent with the experimental results~\cite{anna}, which
found a weak first-order transition from the paramagnetic to the $C_{4}$
phase.

As for the location of the crossing point between the $T_{c}(x)$
and $T_{N}(x)$ lines, it can in principle be in the range where SDW
order is either $C_{2}$ or $C_{4}$, depending on several input parameters
of the model. In the experiments of Ref.~\cite{anna}, the crossing
point happens in the range of $C_{4}$ order. We use this experimental
result as an input and show that, in this situation, there must be
two discontinuities in $T_{c}(x)$ in the region of coexistence with
SDW order, consistent with the experimental findings (iii) and (iv)
discussed in the Introduction.

\textit{Discontinuity of $T_{c}$ at the onset of coexistence with
SDW.}~~~We first consider how $T_{c}$ evolves once the SC transition
line crosses the line of the first-order SDW transition into the $C_{4}$
phase. To achieve this, we write the GL model for coupled SDW and
$s^{+-}$ SC order parameters as \cite{Fernandes10,fern_schm,Vorontsov09}
\begin{equation}
\begin{aligned}F= & \frac{a}{2}\fvec M^{2}+\frac{u}{4}\fvec M^{4}+\frac{\gamma}{6}\fvec M^{6}+\frac{\alpha_{s}}{2}\left|\Delta_{s\pm}\right|^{2}\\
 & +c\left|\Delta_{s\pm}\right|^{2}\fvec M^{2}+\frac{\beta_{s}}{4}\left|\Delta_{s\pm}\right|^{4}\ .
\end{aligned}
\label{Eqn:CoeF}
\end{equation}
where $\alpha_{s}=a_{s}(T-T_{c})$ ($a_{s}>0$) and $M^{2}={\bf M}_{1}^{2}+{\bf M}_{2}^{2}=2{\bf M}_{1}^{2}$.
As explained above, we have $u<0$ and $\gamma>0$, in which case
the SDW transition into the $C_{4}$ phase is first-order. It occurs
at $a=\frac{3u^{2}}{16\gamma}$, and $M^{2}$ jumps from zero to $M_{0}^{2}=-\frac{3u}{4\gamma}$.
An elementary analysis shows that a jump of $M^{2}$ at the SDW transition
gives rise to a discontinuity in $T_{c}$ as $\alpha_{s}$ is renormalized
to $\tilde{\alpha}_{s}=\alpha_{s}+2cM_{0}^{2}$. Hence
\begin{equation}
\delta T_{c}=\frac{3c}{2}\frac{u}{\gamma a_{s}}\ .
\end{equation}
The remaining issue is whether $\delta T_{c}$ is positive or negative.
Because $u<0$ and $\gamma>0$, $\mathrm{sign}\left(\delta T_{c}\right)=-\mathrm{sign}\left(c\right)$.
We computed the coefficient $c$ in terms of parameters of the underlying
fermionic model and found that $c$ is \textit{positive} (the details
of the computations are presented in SI). Therefore, $\delta T_{c}$
is negative, implying that the superconducting transition temperature
jumps down upon entering the coexistence phase with SDW (see Fig.
\ref{Fig:SCSDWPhase}). A negative jump $\delta T_{c}$ is consistent
with the experimental observation (iii) outlined in the Introduction.

\textit{Discontinuity of $T_{c}$ at the boundary between the $C_{2}$
and the $C_{4}$ phases.}~~~ Finally, the experiment reveals that
$T_{c}$ is again discontinuous inside the SDW phase~\cite{anna},
when the SDW order switches from $C_{4}$ back to $C_{2}$ as doping
decreases. Although the $C_{2}-C_{4}$ transition is first order,
the coupling between $\left|\Delta_{s\pm}\right|^{2}$ and $\fvec M^{2}$
cannot explain this discontinuity because $\fvec M^{2}=\fvec M_{1}^{2}+\fvec M_{2}^{2}$
is continuous across the $C_{2}-C_{4}$ SDW phase transition. A more
careful analysis, however, reveals that the $c$ term in the free
energy (\ref{Eqn:CoeF}) arises from the combination of three distinct
microscopic couplings between the magnetic order parameters and the
gap functions at the hole pocket $h$ and the electron pockets $e_{1}$
and $e_{2}$: $c_{hh}\left|\Delta_{h}\right|^{2}\sum_{i}\fvec M_{i}^{2}$,
$c_{ee}\sum_{i}\fvec M_{i}^{2}\left|\Delta_{e_{i}}\right|^{2}$, and
$c_{he}\sum_{i}\fvec M_{i}^{2}\left(\Delta_{h}\Delta_{e_{i}}^{*}+\Delta_{e_{i}}\Delta_{h}^{*}\right)$.
By symmetry, these three superconducting gaps can be equivalently
recast in terms of an $s^{++}$, an $s^{+-}$, and a $d$-wave gap,
as explained in the SI. Neglecting the $s^{++}$ component, which
does not distinguish between the $C_{4}$ and $C_{2}$ phases, we
write the free-energy as
\begin{equation}
\begin{aligned}F= & \frac{a}{2}\fvec M^{2}+\frac{u}{4}\fvec M^{4}+\frac{\gamma}{6}\fvec M^{6}+\frac{\alpha_{s}}{2}\left|\Delta_{s\pm}\right|^{2}+\frac{\alpha_{d}}{2}\left|\Delta_{d}\right|^{2}\\
 & +c_{s}\left|\Delta_{s\pm}\right|^{2}\fvec M^{2}+c_{d}\left|\Delta_{d}\right|^{2}\fvec M^{2}\\
 & +c_{sd}\left(\Delta_{s\pm}^{*}\Delta_{d}+\mathrm{h.c.}\right)\left(\fvec M_{1}^{2}-\fvec M_{2}^{2}\right)+...\ ,
\end{aligned}
\label{eq_F_sd}
\end{equation}
The last term shows that the simultaneous presence of $s^{+-}$ superconductivity
and $C_{2}$ SDW order generates a $d$-wave component of the SC order
parameter \cite{Fernandes_Millis}, even though the leading instability
is not towards a $d$-wave SC phase -- i.e. $\alpha_{d}=a_{d}(T-T_{d})
\approx a_{d}(T_{c}-T_{d})>0$ in Eq. (\ref{eq_F_sd}). In more general terms, once the $C_{4}$
symmetry is broken down to $C_{2}$, the $s-$wave and $d-$wave SC
order parameters no longer belong to different irreducible representations
of the point group symmetry and the presence of one causes the appearance
of the other.

We can now analyze the behavior of $T_{c}$ in the coexistence phase
with SDW. If the SDW is the $C_{4}$ phase, where $\fvec M_{1}^{2}=\fvec M_{2}^{2}$,
the last term in Eq. (\ref{eq_F_sd}) is irrelevant, and the SC transition
temperature is determined by ${\tilde{\alpha}}_{s}=\alpha_{s}+2c_{s}\fvec M^{2}=0$,
i.e.
\[
T_{c}^{(C_{4})}=
T_{c}-\frac{2c_{s}}{a_{s}}\fvec M^{2}
\]
If the SDW is the $C_{2}$ phase, the quadratic part of the SC GL
free energy is given by:
\begin{equation}
F_{\mathrm{SC}}=\half\left(\begin{array}{c}
\Delta_{s\pm}\\
\Delta_{d}
\end{array}\right)^{T}\left(\begin{array}{cc}
{\tilde{\alpha}}_{s} & 2c_{sd}\varphi\\
2c_{sd}\varphi & {\tilde{\alpha}}_{d}
\end{array}\right)\left(\begin{array}{c}
\Delta_{s\pm}\\
\Delta_{d}
\end{array}\right)\ ,\label{Eqn:C2SDWSDF}
\end{equation}
where ${\tilde{\alpha}}_{s}=\alpha_{s}+2c_{s}\fvec M^{2}$, ${\tilde{\alpha}}_{d}=\alpha_{d}+2c_{d}\fvec M^{2}$,
and $\varphi=\fvec M_{1}^{2}-\fvec M_{2}^{2}$. Diagonalizing the
matrix, we find that the superconducting $T_{c}$ in the $C_{2}$
phase is given by ${\tilde{\alpha}}_{s}{\tilde{\alpha}}_{d}=(2c_{cd}\varphi)^{2}$,
hence
\begin{equation}
T_{c}^{\left(C_{2}\right)}=T_{c}^{\left(C_{4}\right)}+\frac{4c_{cs}^{2}\varphi^{2}}{a_{s}a_{d}\left(
T_{c}-T_{d}\right)}
\end{equation}
The key point here is that even though $\fvec M^{2}$ changes continuously
across the $C_{4}\rightarrow C_{2}$ transition, the quantity $\varphi=\fvec M_{1}^{2}-\fvec M_{2}^{2}$
jumps from $\varphi=0$ in the $C_{4}$ phase to $\varphi=\pm\fvec M^{2}$
in the $C_{2}$ phase. As a result, $T_{c}$ jumps \textit{up} once
the system moves from $C_{4}$ to $C_{2}$ SDW order inside the SDW-SC
coexistence state. This is consistent with the experimental result
(iv) discussed in the Introduction~\cite{anna}. Note that the near
degeneracy between the $s^{+-}$ and the $d$-wave states, as attested
by Raman scattering experiments \cite{raman_mode1,raman_mode2} in
optimally doped Ba$_{1-x}$K$_{x}$Fe$_{2}$As$_{2}$ implies that
the $T_{c}$ and $T_{d}$ values are close, causing a visible jump
in $T_{c}$.

\textit{Conclusions.}~~~ In this communication we analyzed the
structure of the SDW order arising from an itinerant fermionic model
in doped iron-pnictides and its impact on the superconducting $T_{c}$
in the coexistence phase with magnetism. We found that stripe magnetic
order does not occur at all doping/temperatures where a magnetic instability
is present -- in particular, there is a narrow doping/temperature
range located near the magnetic transition line $T_{N}(x)$ where
the SDW order preserves the $C_{4}$ lattice rotational symmetry.
We argued that, as the SC transition line crosses the SDW transition
line, the superconducting $T_{c}$ has two discontinuities -- it jumps
to a smaller value upon entering the coexistence region with $C_{4}$
SDW, and it jumps to a larger value inside the SDW state, when it
crosses the boundary between $C_{4}$ and $C_{2}$ SDW orders. The
resulting phase diagram, schematically shown in Fig. \ref{Fig:SCSDWPhase},
is almost identical to the experimental phase diagram of the K-doped
122 material~\cite{anna}. We view the agreement between theory and
experiment, even in their fine details, as a strong indication that
the itinerant approach is adequate to describe the physics of weakly/moderately
doped Fe-pnictides.

We thank A. E. B\"ohmer, F. Hardy, and C. Meingast, for useful discussions
and for sharing their data with us prior to publication. This work
was supported by the Office of Basic Energy Sciences U. S. Department
of Energy under awards numbers DE-SC0012336 (XW, JK, and RMF) and
DE-FG02-ER46900 (AVC).

\newpage

\widetext
\vspace{0.5cm}
\begin{center}
\textbf{\large Supplementary for ``Interplay between tetragonal magnetic order, stripe magnetism, and
superconductivity in iron-based materials''}
\end{center}
\setcounter{equation}{0}
\setcounter{figure}{0}
\setcounter{table}{0}
\makeatletter
\renewcommand{\theequation}{S\arabic{equation}}
\renewcommand{\thefigure}{S\arabic{figure}}
\renewcommand{\bibnumfmt}[1]{[S#1]}
\renewcommand{\citenumfont}[1]{S#1}


\section{Derivation of the free-energy in the absence of superconductivity}

We first discuss the free-energy of the pure magnetic system, which
is given, up to sixth-order in the magnetic order parameters, by:

\begin{equation}
\begin{aligned}F(\fvec M_{i})= & \frac{a}{2}\left(\fvec M_{1}^{2}+\fvec M_{2}^{2}\right)+\frac{u}{4}\left(\fvec M_{1}^{2}+\fvec M_{2}^{2}\right)^{2}-\frac{g}{4}\left(\fvec M_{1}^{2}-\fvec M_{2}^{2}\right)^{2}+w\left(\fvec M_{1}\cdot\fvec M_{2}\right)^{2}\\
 & -\frac{v}{6}\left(\fvec M_{1}^{2}-\fvec M_{2}^{2}\right)^{2}\left(\fvec M_{1}^{2}+\fvec M_{2}^{2}\right)+\frac{\gamma}{6}\left(\fvec M_{1}^{2}+\fvec M_{2}^{2}\right)^{3}+\frac{\lambda}{6}\left(\fvec M_{1}\cdot\fvec M_{2}\right)^{2}\left(\fvec M_{1}^{2}+\fvec M_{2}^{2}\right)
\end{aligned}
\label{Eqn:AllF}
\end{equation}

To calculate the coefficients in Eqn.~(\ref{Eqn:AllF}), we follow
Ref.~\cite{Fernandes12_S} and start from the Hamiltonian $H=H_{0}+H_{\mathrm{int}}$,
where $H_{0}$ is the 3-band non-interacting Hamiltonian discussed
in Eq. (1) of the main text, and $H_{\mathrm{int}}$ contains the
projections of all interactions into the SDW channel. We decouple
these interaction terms by Hubbard-Stratonovich transformations and
introduce the fields $\fvec M_{i}(\fvec q)=U_{SDW}\sum_{k}c_{\fvec k+\fvec q,h}^{\dagger}\frac{\fvec\sigma}{2}c_{\fvec k+\fvec Q_{i},e_{i}}$,
whose mean values are the magnetic order parameters. The interacting
Hamiltonian becomes:
\begin{equation}
H_{SDW}=\sum_{\mathbf{k},i}\left[\fvec M_{i}\cdot\left(c_{\fvec k+\fvec q,h}^{\dagger}\frac{\fvec\sigma}{2}c_{\fvec k+\fvec Q_{i},e_{i}}+\mathrm{h.c.}\right)+\frac{\fvec{M}_{i}^{2}}{2U_{SDW}}\right]
\end{equation}

To simplify the notation, it is convenient to write $H_{0}$ and $H_{SDW}$
in the basis of a Nambu spinor $\psi$,
\begin{equation}
H_{SDW}=\sum_{\mathbf{k}}\psi_{\mathbf{k}}^{\dag}\hat{H}_{1,\mathbf{k}}(\fvec{M}_{i})\psi_{\mathbf{k}}\qquad H_{0}=\sum_{\mathbf{k}}\psi_{\mathbf{k}}^{\dag}\hat{H}_{0}\psi_{\mathbf{k}}\ .
\end{equation}

Because the Hamiltonian is now quadratic in the fermions, they can
be integrated out in the partition function, and the partition function
can be expressed as the functional integral over ${\bf M}_{i}$ fields
\begin{align}
Z & \propto\int\mathcal{D}\boldsymbol{M}_{i}\det\left(\hat{G}_{0}^{-1}-\hat{G}_{1}\right)\exp\left(-\int\left(\frac{\fvec{M}_{1}^{2}+\fvec{M}_{2}^{2}}{2U_{SDW}}\right)\right)\nonumber \\
 & =\det(\hat{G}_{0}^{-1})\int\mathcal{D}\boldsymbol{M}_{i}\det\Big(1-\hat{G}_{0}\hat{H}_{1}\Big)\exp\left(-\int\left(\frac{\fvec{M}_{1}^{2}+\fvec{M}_{2}^{2}}{2U_{SDW}}\right)\right)
\end{align}
Here $\hat{G}_{0}$ is the Green's function of the free fermions,
$\hat{G}_{0}=i\omega_{n}\hat{I}-\hat{H}_{0}$. Expanding the action
in powers of the order parameters $\fvec M_{i}$ we obtain
\begin{align}
Z & =\int\mathcal{D}\boldsymbol{M}_{i}\exp(-S_{eff})\\
S_{eff} & =-\ln\det(1-\hat{G}_{0}\hat{H}_{1})+\frac{\fvec{M}_{1}^{2}+\fvec{M}_{2}^{2}}{2U_{SDW}}\nonumber \\
 & =-\mathrm{tr}\ln(1-\hat{G}_{0}\hat{H}_{1})+\frac{\fvec{M}_{1}^{2}+\fvec{M}_{2}^{2}}{2U_{SDW}}=\sum_{n}\frac{1}{n}\mathrm{tr}\left(\hat{G}_{0}\hat{H}_{1}\right)^{n}+\frac{\fvec{M}_{1}^{2}+\fvec{M}_{2}^{2}}{2U_{SDW}}
\end{align}
It is now straightforward to derive the coefficients of Eq. (\ref{Eqn:AllF}).
For the quartic coefficients, we find $w=0$ and:
\begin{equation}
\begin{aligned}u & =A+B\ ,\quad g=B-A\\
A & =\int_{k}\left(G_{h}(\bk)\right)^{2}\left(G_{e_{1}}(\bk+\fvec Q_{1})\right)^{2}=\int_{k}\left(\frac{1}{i\omega+\epsilon}\right)^{2}\left(\frac{1}{i\omega-\epsilon-\delta_{\mu}-\delta_{m}\cos2\theta}\right)^{2}\ ,\\
B & =\int_{k}\left(G_{h}(\bk)\right)^{2}G_{e_{1}}(\bk+\fvec Q_{1})G_{e_{2}}(\bk+\fvec Q_{2})=\int_{k}\left(\frac{1}{i\omega+\epsilon}\right)^{2}\frac{1}{i\omega-\epsilon-\delta_{\mu}-\delta_{m}\cos2\theta}\frac{1}{i\omega-\epsilon-\delta_{\mu}+\delta_{m}\cos2\theta}\ .
\end{aligned}
\label{Eqn:QCoe}
\end{equation}
Here, $G_{a}(\fvec k)$ is the free-fermion Green's function for pocket
$a$, $G_{a}^{-1}(\bk)=i\omega_{n}-\epsilon_{a}(\bk)$, and $\int_{k}\rightarrow T\sum_{n}\int\dfrac{\rmd^{d}k}{(2\pi)^{d}}$,
with Matsubara frequency $\omega_{n}=(2n+1)\pi T$. After integrating
over the momentum we obtain:
\begin{equation}
\begin{aligned}A= & N_{f}T\pi\sum_{n=0}^{\infty}\Im\int_{\theta}\frac{1}{\left(i\omega_{n}+\dfrac{\delta_{\mu}}{2}+\dfrac{\delta_{m}}{2}\cos2\theta\right)^{3}}\\
B= & N_{f}T\pi\sum_{n=0}^{\infty}\Im\int_{\theta}\frac{i\omega_{n}+\dfrac{\delta_{\mu}}{2}}{\left(\left(i\omega_{n}+\dfrac{\delta_{\mu}}{2}\right)^{2}-\left(\dfrac{\delta_{m}}{2}\cos2\theta\right)^{2}\right)^{2}}
\end{aligned}
\end{equation}
with $\int_{\theta}=\int\frac{\rmd\theta}{2\pi}$ and $N_{f}$ is
the density of states. For the sixth-order coefficients, we obtain
$\lambda=0$ and:
\begin{equation}
\begin{aligned}\gamma & =C+3D\ ,\qquad v=3(D-C)\\
C= & \int_{k}\left(G_{h}(\bk)\right)^{3}\left(G_{e_{1}}(\bk+\fvec Q_{1})\right)^{3}=N_{f}T\pi\frac{3}{4}\sum_{n=0}^{\infty}\Im\int_{\theta}\frac{1}{\left(i\omega_{n}+\dfrac{\delta_{\mu}}{2}+\dfrac{\delta_{m}}{2}\cos2\theta\right)^{5}}\quad,\\
D= & \int_{k}\left(G_{h}(\bk)\right)^{3}\left(G_{e_{1}}(\bk+\fvec Q_{1})\right)^{2}G_{e_{2}}(\bk+\fvec Q_{2})\\
= & \frac{N_{f}T\pi}{8}\sum_{n=0}^{\infty}\Im\int_{\theta}\frac{3}{\left(i\omega_{n}+\dfrac{\delta_{\mu}}{2}+\dfrac{\delta_{m}}{2}\cos2\theta\right)^{4}\left(i\omega_{n}+\dfrac{\delta_{\mu}}{2}-\dfrac{\delta_{m}}{2}\cos2\theta\right)}\\
 & +\frac{2}{\left(i\omega_{n}+\dfrac{\delta_{\mu}}{2}+\dfrac{\delta_{m}}{2}\cos2\theta\right)^{3}\left(i\omega_{n}+\dfrac{\delta_{\mu}}{2}-\dfrac{\delta_{m}}{2}\cos2\theta\right)^{2}}\\
 & +\frac{1}{\left(i\omega_{n}+\dfrac{\delta_{\mu}}{2}+\dfrac{\delta_{m}}{2}\cos2\theta\right)^{2}\left(i\omega_{n}+\dfrac{\delta_{\mu}}{2}-\dfrac{\delta_{m}}{2}\cos2\theta\right)^{3}}\ .\label{Eqn:SCoe}
\end{aligned}
\end{equation}

We can now evaluate $g$ numerically at any temperature and analytically
at high temperatures and at $T=0$. For $T\gg\delta_{\mu}$ and $T\gg\delta_{m}$
we obtain:
\begin{align}
A\approx & N_{f}T\pi\Im\int_{\theta}\sum_{n=0}^{\infty}\frac{1}{(i\omega_{n})^{3}}\left(1-3\frac{\delta_{\mu}+\delta_{m}\cos2\theta}{2i\omega_{n}}+6\left(\frac{\delta_{\mu}+\delta_{m}\cos2\theta}{2i\omega_{n}}\right)^{2}\right)\nonumber \\
B\approx & N_{f}T\pi\Im\int_{\theta}\sum_{n=0}^{\infty}\frac{1}{(i\omega_{n})^{3}}\left(1-3\frac{\delta_{\mu}}{2i\omega_{n}}+2\left(\frac{\delta_{m}\cos2\theta}{2i\omega_{n}}\right)^{2}+6\left(\frac{\delta_{\mu}}{2i\omega_{n}}\right)^{2}\right)\\
g= & B-A\approx-N_{f}T\pi\Im\int_{\theta}\sum_{n=0}^{\infty}\frac{(\delta_{m}\cos2\theta)^{2}}{(i\omega_{n})^{5}}=\frac{\pi TN_{f}\delta_{m}^{2}}{2(2\pi T)^{5}}\sum_{n=0}^{\infty}\frac{1}{(n+1/2)^{5}}=\frac{31\zeta(5)N_{f}\delta_{m}^{2}}{64\pi^{4}T^{4}}\ .\nonumber
\end{align}

At $T=0$ we have $T\sum_{n}\rightarrow\int\frac{\rmd\omega}{2\pi}$.
To regularize the integral, we add the lifetime $\mathrm{sign}(\omega_{n})/(2\tau)$
to the fermion propagator and take the limit $\tau\rightarrow\infty$
after we compute $g$. We obtain:
\begin{align}
A= & -N_{f}\int_{\theta}\Re\left(\frac{1}{\delta_{\mu}+\delta_{m}\cos2\theta+i/\tau}\right)^{2}\nonumber \\
B= & -N_{f}\int_{\theta}\Re\frac{1}{(\delta_{\mu}+\delta_{m}\cos2\theta+i/\tau)(\delta_{\mu}-\delta_{m}\cos2\theta+i/\tau)}\\
g= & B-A=\frac{N_{f}}{2}\int_{\theta}\Re\left(\frac{1}{\delta_{\mu}+\delta_{m}\cos2\theta+i/\tau}-\frac{1}{\delta_{\mu}-\delta_{m}\cos2\theta+i/\tau}\right)^{2}
\label{eq_g}
\end{align}

At $T_{N}\rightarrow0$, $\left|\delta_{\mu}\right|>\delta_{m}$ (see
Ref. \cite{Vorontsov09_S}). In this case we can safely set $\tau=\infty$
in the integrals in (\ref{eq_g}) and obtain:
\begin{equation}
g=\frac{\delta_{m}^{2}N_{f}}{|\delta_{\mu}|(\delta_{\mu}^{2}-\delta_{m}^{2})^{3/2}}\geq0\ .
\end{equation}

If this condition was not satisfied, i.e. if $\left|\delta_{\mu}\right|\leq\delta_{m}$,
$\tau^{-1}$ must remain finite to avoid a divergence in $g$. Analytical
evaluation of the integral then reveals that $g\propto1/\tau$ and
\begin{equation}
g>0\mbox{ if }|\delta_{\mu}|<\delta_{m}/2\ .
\end{equation}

\begin{figure}[htbp]
\begin{centering}
\includegraphics[width=0.45\columnwidth]{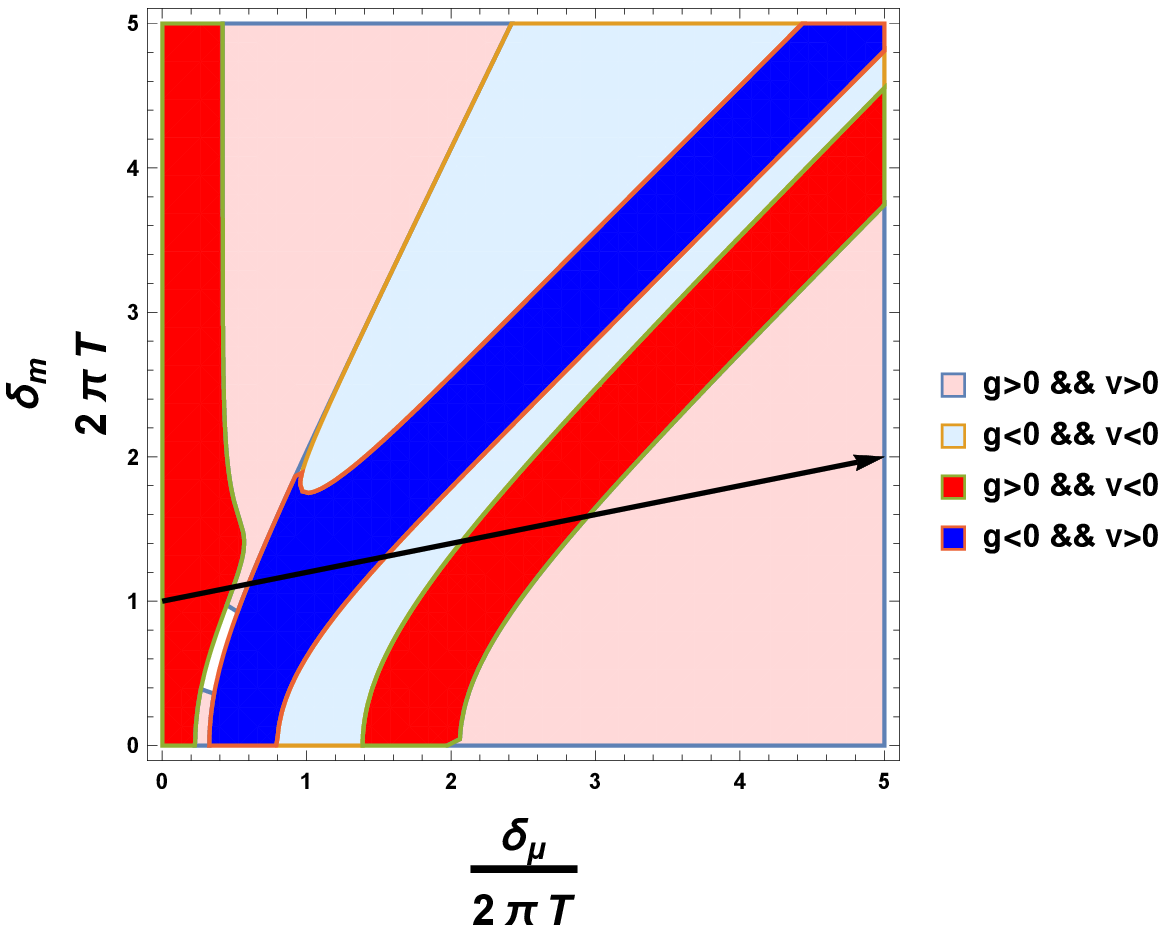}
\par\end{centering}

\bigskip{}

\centering
\includegraphics[width=0.45\columnwidth]{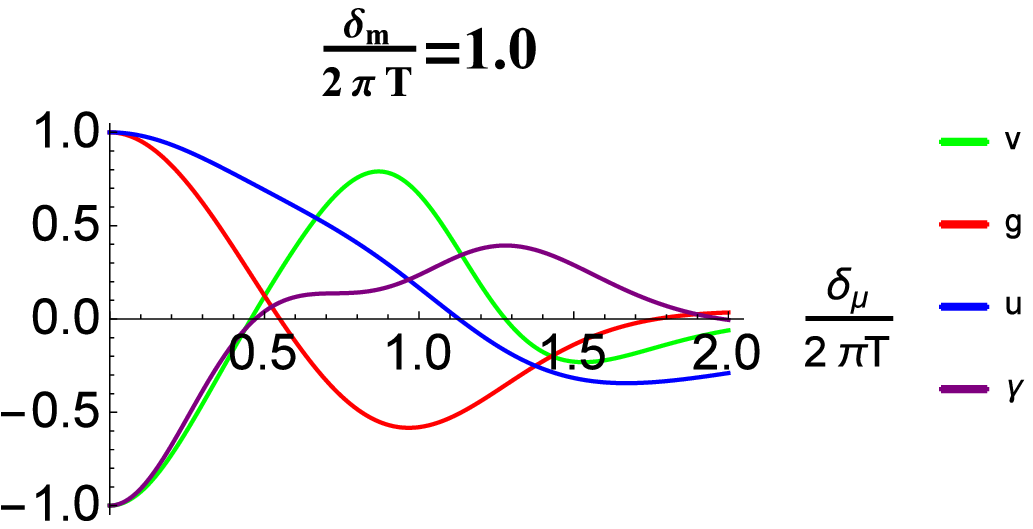}
\caption{(\textbf{upper panel}) The signs of $g$ and $v$ as a function of
$\delta_{\mu}/(2\pi T)$ and $\delta_{m}/(2\pi T)$. The arrow represents
schematically a path along the magnetic transition line $T=T_{N}(x)$,
in which $\delta_{m}$ is a constant and $\delta_{\mu}$ increases
with doping. The arrow points towards larger doping. (\textbf{lower
panel}) The quartic ($u$, $g$) and sixth-order ($v$, $\gamma$)
SDW coefficients as a function of $\delta_{\mu}/(2\pi T)$ when $\delta_{m}/(2\pi T)$
is fixed to be $1.0$. Note that these coefficients are normalized
by their values at $\delta_{\mu}=0$. }
\label{Fig:SDW_suppl}
\end{figure}

For intermediate temperatures we evaluate $g$ -- and also $v$ --
numerically. The result is shown in Fig. \ref{Fig:SDW_suppl} in the
$\delta_{\mu}/\left(2\pi T\right)$, $\delta_{m}/\left(2\pi T\right)$
plane. We see that, as temperature decreases and one tracks the magnetic
transition line $T_{N}\left(x\right)$, there is a sequence of changes
from $g>0$ to $g<0$ and then back to $g>0$ (the arrow in the plot
represents a schematic path along the magnetic transition line). In
Fig. 1 in the main text, we plotted a cut for the fixed value $\delta_{m}/\left(2\pi T\right)=1$,
which is representative of this behavior. We plot again this cut in
Fig. \ref{Fig:SDW_suppl}, together with the behavior of $u$ and
$\gamma$. As discussed in the main text, there is a regime inside
the $C_{4}$ phase where $u<0$ and $\gamma>0$, which implies that
the SDW transition is first order.

\section{Derivation of the Free energy in the presence of superconductivity}

The free-energy in the presence of both SC and SDW degrees of freedom
is given by:

\begin{equation}
\begin{aligned}F(\fvec M_{i},\Delta_{m})= & F(\fvec M_{i})+\frac{\alpha_{s}}{2}\left|\Delta_{s\pm}\right|^{2}+\frac{\alpha_{d}}{2}\left|\Delta_{d}\right|^{2}+\frac{\beta_{s}}{4}\left|\Delta_{s\pm}\right|^{4}+\frac{\beta_{d}}{4}\left|\Delta_{d}\right|^{4}\\
 & +c_{s}\left|\Delta_{s\pm}\right|^{2}\left(\fvec M_{1}^{2}+\fvec M_{2}^{2}\right)+c_{d}\left|\Delta_{d}\right|^{2}\left(\fvec M_{1}^{2}+\fvec M_{2}^{2}\right)+c_{sd}\left(\Delta_{s\pm}^{*}\Delta_{d}+\mathrm{h.c.}\right)\left(\fvec M_{1}^{2}-\fvec M_{2}^{2}\right)
\end{aligned}
\label{F_SC_SDW}
\end{equation}

To derive the SC coefficients, we need to account also for the interactions
in $H_{\mathrm{int}}$ that promote superconductivity. Following Ref.
\cite{Fernandes_Millis_S}, we express them in terms of the inter-band
pairing interactions $U_{eh}>0$ (between the hole pocket and either
of the electron pockets) and $U_{ee}>0$ (between the two electron
pockets). Introducing the gap function at each pocket, $\Delta_{i}=\sum_{\fvec k,j}U_{ij}c_{-\fvec k,j\downarrow}c_{\fvec k,j\uparrow}$,
we write the interacting SC Hamiltonian as
\begin{equation}
H_{SC}=\sum_{\fvec k,ij}\left[\left(\Delta_{i}c_{\fvec k,i\uparrow}^{\dagger}c_{-\fvec k,i\downarrow}^{\dagger}+\mathrm{h.c.}\right)\delta_{ij}+\frac{1}{2}\Delta_{i}U_{ij}^{-1}\Delta_{j}^{*}\right]
\end{equation}

To proceed, we repeat the same steps as in the previous section, but
with $H_{SDW}+H_{SC}$ instead of $H_{SDW}$. It is convenient to
switch to a parametrization of the superconducting order parameters
in terms of the irreducible representations $A_{1g}$ ($s_{\pm}$
and $s_{++}$ states) and $B_{1g}$ ($d$-wave state) \cite{Fernandes_Millis_S}:
\begin{equation}
\left(\begin{array}{c}
\Delta_{h}\\
\Delta_{e_{1}}\\
\Delta_{e_{2}}
\end{array}\right)=\left(\begin{array}{ccc}
\sin\phi & -\cos\phi & 0\\
\frac{\cos\phi}{\sqrt{2}} & \frac{\sin\phi}{\sqrt{2}} & -\frac{1}{\sqrt{2}}\\
\frac{\cos\phi}{\sqrt{2}} & \frac{\sin\phi}{\sqrt{2}} & \frac{1}{\sqrt{2}}
\end{array}\right)\left(\begin{array}{c}
\Delta_{s_{++}}\\
\Delta_{s_{\pm}}\\
\Delta_{d}
\end{array}\right)
\end{equation}
where the parameter $\phi$ depends on the ratio of the pairing interactions:
\begin{equation}
\tan\phi=\frac{\sqrt{8U_{eh}^{2}+U_{ee}^{2}}-U_{ee}}{2\sqrt{2}U_{eh}}
\end{equation}

Note that the SC gaps in the $s_{\pm}$ state have different signs
in the hole and electron pockets:
\begin{equation}
\Delta_{e_{1}}=\Delta_{e_{2}}=-\frac{\tan\phi}{\sqrt{2}}\Delta_{h}\ .
\end{equation}

In the $s_{++}$ state, on the other hand, all SC gaps have the same
sign:
\begin{equation}
\Delta_{e_{1}}=\Delta_{e_{2}}=\frac{\cot\phi}{\sqrt{2}}\Delta_{h}\ .
\end{equation}

In the d-wave state, the gap in the hole pocket averages to $0$ and
the gaps in the two electron pockets have opposite signs.

Applying this transformation and ignoring the $\Delta_{s_{++}}$ contributions
we obtain:
\begin{equation}
\begin{split}F(\fvec M_{i},\Delta_{m}) & =F(\fvec M_{i})+\frac{1}{2}\left(a_{sh}\cos^{2}\phi+a_{se}\sin^{2}\phi\right)|\Delta_{s_{+-}}|^{2}+\frac{1}{2}a_{se}|\Delta_{d}|^{2}\\
 & +\frac{1}{4}\left(u_{sh}\cos^{4}\phi+\frac{1}{2}u_{se}\sin^{4}\phi\right)|\Delta_{s_{+-}}|^{4}+\frac{1}{8}u_{se}|\Delta_{d}|^{4}+\frac{1}{4}u_{se}\sin^{2}\phi|\Delta_{s_{+-}}|^{2}|\Delta_{d}|^{2}\left(1+2\cos^{2}\theta\right)\\
 & +\left[\frac{1}{2}\left(\gamma_{h}\cos^{2}\phi+\frac{1}{2}\gamma_{e}\sin^{2}\phi-\frac{1}{\sqrt{2}}\gamma_{he}\sin2\phi\right)|\Delta_{s_{+-}}|^{2}+\frac{1}{4}\gamma_{e}|\Delta_{d}|^{2}\right]\left(\fvec M_{1}^{2}+\fvec M_{2}^{2}\right)\\
 & +\frac{1}{2}\left(\frac{1}{\sqrt{2}}\gamma_{he}\cos\phi-\frac{1}{2}\gamma_{e}\sin\phi\right)\left(\Delta_{s\pm}^{*}\Delta_{d}+\Delta_{s\pm}\Delta_{d}^{*}\right)\left(\fvec M_{1}^{2}-\fvec M_{2}^{2}\right)
\end{split}
\end{equation}

The coupling constants describing the interplay between SC and SDW
are given by:
\begin{equation}
\begin{split}\gamma_{h} & =16\sum_{k}G_{hk}^{2}\tilde{G}_{hk}G_{e_{1}k}\\
\gamma_{e} & =16\sum_{k}G_{e_{1}k}^{2}\tilde{G}_{e_{1}k}G_{hk}\\
\gamma_{he} & =8\sum_{k}G_{hk}\tilde{G}_{hk}G_{e_{1}k}\tilde{G}_{e_{1}k}
\end{split}
\end{equation}

Here $\tilde{G}$ is the Green function for the hole states, $\tilde{G}=(-G)^{*}$.
The coefficients of Eq. (\ref{F_SC_SDW}) are given by:

\begin{align}
c_{s} & =\frac{1}{2}\left(\gamma_{h}\cos^{2}\phi+\frac{1}{2}\gamma_{e}\sin^{2}\phi-\frac{1}{\sqrt{2}}\gamma_{he}\sin2\phi\right)\nonumber \\
c_{d} & =\frac{1}{4}\gamma_{e}\nonumber \\
c_{sd} & =\frac{1}{2}\left(\frac{1}{\sqrt{2}}\gamma_{he}\cos\phi-\frac{1}{2}\gamma_{e}\sin\phi\right)
\end{align}

\begin{figure}[htbp]
\centering
\includegraphics[scale=0.7]{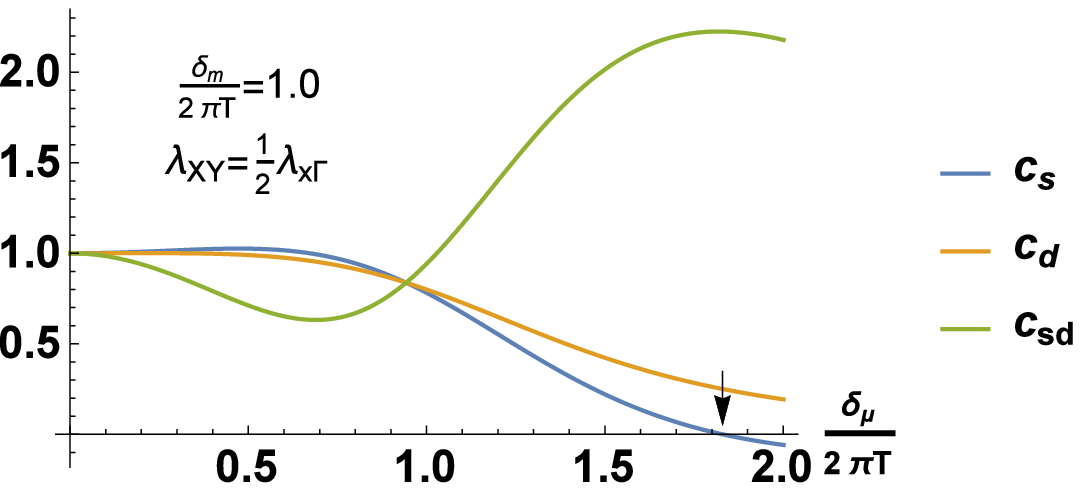}
\caption{Coupling constants $c_{s}$, $c_{d}$, and $c_{sd}$, describing the
interplay between the SDW and SC order parameters, as function of
$\delta_{\mu}/\left(2\pi T\right)$ for a fixed $\delta_{m}/\left(2\pi T\right)=1$.
The couplings constants are normalized to their values at zero doping
($\delta_{\mu}=0$). The arrow shows the value where $c_{s}$ changes
sign, which according to Fig. \ref{Fig:SDW_suppl} happens very close
to the $C_{4}\rightarrow C_{2}$ transition point.}
\label{Fig:SCSDWCoupling_Suppl}
\end{figure}

At perfect nesting, where $\delta_{\mu}=\delta_{2}=0$, we find that
$\gamma_{h}=\gamma_{e}=2\gamma_{he}=\frac{14\zeta(3)N_{F}}{\pi^{2}T^{2}}$.
In the limit where $U_{he}\gg U_{ee}$, $\phi\rightarrow\pi/4$ and
we find that $c_{s}=\left(3-\sqrt{2}\right)\gamma_{h}/8>0$. The positive
sign of $c_{s}$ is evidence of the competition between $s_{\pm}$
SC and SDW. Of course, the value of $c_{s}$ depends on the parameter
$\phi$, which is sensitive to the ratio between the pairing interactions
$U_{he}$ and $U_{ee}$. In particular, for $U_{he}>U_{ee}$, the
ground state is $s_{\pm}$ whereas for $U_{he}<U_{ee}$, the ground
state is $d$-wave. Thus, to describe the Ba$_{1-x}$K$_{x}$Fe$_{2}$As$_{2}$
system, for which the $s_{\pm}$ and $d$-wave states are close in
energy, we take $U_{ee}$ to be somewhat below $U_{he}$. For practical
calculations, we used $U_{ee}=U_{he}/2$. In Fig~\ref{Fig:SCSDWCoupling_Suppl}
we show the behavior of the coupling constants $c_{i}$ as functions
of $\delta_{\mu}/\left(2\pi T\right)$ for a fixed $\delta_{m}/\left(2\pi T\right)=1$.
The coupling $c_{s}$ is positive in a wide doping range and, in particular,
in almost the entire doping range where the $C_{4}$ phase exists
(see Fig. \ref{Fig:SDW_suppl}) and where the $T_{c}$ transition
line crosses the $T_{N}$ transition line, according to the experiments
in Ba$_{1-x}$K$_{x}$Fe$_{2}$As$_{2}$ .

\begin{thebibliography}{10}
\bibitem{comm} In principle, the system may develop incommensurate
magnetic order before the transition ecomes first-order ~\cite{Fernandes12}.
We do not consider this possibility as experiments on hole-doped Ba$_{1-x}$K$_{x}$Fe$_{2}$As$_{2}$
indicate that the SDW transition is first-order.
\end{thebibliography}
\end{document}